\documentclass[useAMS]{mn2e}
\usepackage{graphicx}
\usepackage{amsmath}
\usepackage{amssymb}
\usepackage{color}
\usepackage{url}

\newcommand  \acc     {\ifmmode {\rm km\,s}^{-2} \else km\,s$^{-2}$\fi}

\newcommand  \ergs     {\ifmmode {\rm ergs\,s}^{-1} \else ergs s$^{-1}$\fi}
\newcommand  \ergcms   {\ifmmode {\rm erg~cm}^{-2}\,{\rm s}^{-1}
                        \else erg~cm$^{-2}$\,s$^{-1}$\fi}
\newcommand  \ergcmsA  {\ifmmode{\rm erg\,cm}^{-2}\,{\rm s}^{-1}\,{\rm\AA}^{-1}
                        \else ergs\,cm$^{-2}$\,s$^{-1}$\,\AA$^{-1}$\fi}
\newcommand  \ergcmsHz {\ifmmode{\rm ergs\,cm}^{-2}\,{\rm s}^{-1}\,{\rm Hz}^{-1}
                        \else ergs\,cm$^{-2}$\,s$^{-1}$\,Hz$^{-1}$\fi}
\newcommand  \phcms    {\ifmmode {\rm ph\,cm}^{-2}\,{\rm s}^{-1}
                        \else ph\,cm$^{-2}$\,s$^{-1}$\fi}
\newcommand  \phcmsA   {\ifmmode {\rm ph\,cm}^{-2}\,{\rm s}^{-1}\,{\rm\AA}^{-1}
                        \else ph\,cm$^{-2}$\,s$^{-1}$\,\AA$^{-1}$\fi}
\newcommand\aj{{AJ}}% 
\newcommand\apj{{ApJ}}% 
\newcommand\apjl{{ApJ}}% 
\newcommand\apjs{{ApJS}}% 
\newcommand\aap{{A\&A}}% 
\newcommand\mnras{{MNRAS}}% 
\newcommand\pasp{{PASP}}% 
\newcommand\pasa{{PASA}}% 
\newcommand\nat{{Nature}}% 
\newcommand\memsai{{Mem.~Soc.~Astron.~Italiana}}% 
\title[Bio-markers in earths transiting  white dwarfs]
{Detecting bio-markers in habitable-zone earths transiting white dwarfs}
\author[A. Loeb, D. Maoz]
{Abraham Loeb$^{1,2}$, Dan Maoz$^{2}$\\
$^{1}$Institute for Theory and Computation, 
Harvard University, Cambridge, MA 03210, USA\\
$^{2}$School of Physics and Astronomy, Tel-Aviv University, Tel-Aviv 69978,
Israel} \date{\today}

\begin{document}

\maketitle

\label{firstpage}

\begin{abstract}
The characterization of the atmospheres of habitable-zone Earth-mass
exoplanets that transit across main-sequence stars, let alone the
detection of bio-markers in their atmospheres, will be challenging
even with future facilities. It has been noted that white dwarfs (WDs)
have long-lived habitable zones and that a large fraction of WDs may
host planets. We point out that during a transit of an Earth-mass
planet across a WD, the planet's atmospheric transmission spectrum
obtains a much higher contrast over the stellar background compared to
a main-sequence host, because of the small surface area of the WD.
The most prominent bio-marker in the present-day terrestrial
atmosphere, molecular oxygen, is readily detectable in a WD transit
via its A-band absorption at $\sim 0.76 \mu$m.  A potentially
life-sustaining Earth-like planet transiting a WD can be found by
assembling a suitable sample of $\sim 500$~WDs and then surveying them
for transits using small telescopes. If and when a transiting case
 is found, the
O$_2$ absorption in the planetary atmospheric transmission spectrum
would be detectable with the James Webb Space Telescope (JWST) in
about 5 hours of total exposure time, integrated over 160 2-minute
transits. Characterization of the planet atmosphere using other
tracers such as water vapour and CO$_2$  will be considerably
easier. We demonstrate this future discovery space by simulating a
possible transmission spectrum that would be obtained with JWST.
\end{abstract}


%.

%\newpage

\newpage

\section{Introduction}
The discovery of the first transiting exoplanet,
HD209458 (Charbonneau et al. 2000) was quickly followed by a
measurement and characterization of the planet's atmospheric
transmission spectrum (Charbonneau et al. 2002). Spectroscopic
observations of this type provide invaluable probes of planetary
physics, formation, and evolution.  The measurement, however, is
difficult because of the tiny contrast, $\sim 10^{-3}-10^{-4}$,
between the signal (the absorption lines in the light transmitted
through the planet atmosphere) and the background (the unobstructed
light from the host star). In HD209458 the observation was possible
owing to the closeness (and hence brightness, $V=7.7$~mag) of the
star, combined with the stability provided by the {\it Hubble Space
Telescope}. Only a few measurements of this kind have been successful
to date (e.g Tinetti et al. 2007; Redfield et al. 2008; Zhou \&
Bayliss 2012; see also Berta et al. 2012 for a recent null result).
Future prospects for exoplanet transmission spectra have focused on
the capabilities of the {\it James Webb Space
Telescope}\footnote{http://www.jwst.nasa.gov/} (JWST), and on space
mission concepts such as {\it Darwin}
\footnote{http://www.esa.int/Our$_{-}$Activities/Space$_{-}$Science/Darwin$_{-}$overview/}
and the {\it Terrestrial Planet
Finder}\footnote{http://science.nasa.gov/missions/tpf/}.

JWST and other future telescopes may indeed be able to extend
exoplanet atmospheric measurements down to planet masses of a few
earths, particularly in the habitable zones around M-stars (e.g.,
Webb \& Wormleaton 2001; 
Ehrenreich et al.  2006; Beckwith 2008; Kaltenegger \& Traub 2009;
Deming et al. 2009; Rauer et al. 2011; Pall{\'e} et al. 2011; Benneke
\& Seager 2012), though at the price of many tens to hundreds of hours
of total exposure time, integrated over many transits, and only if
such transiting planets exist around nearby M-dwarfs that are bright
enough. Current studies of potentially detectable species in
Earth-like atmospheres have focused on water vapour, methane, CO$_2$,
O$_2$, and ozone in the near- to mid-infrared (IR) part of the
spectrum. The detection of bio-markers that signal the presence of
life on a planet, such as ozone, O$_2$, or even the ``red edge'' of
chlorophyll, is more challenging, and most studies (e.g. von Paris et
al. 2013) have expressed pessimistic prospects for their detection
even with post-JWST space missions currently considered.

On Earth, O$_2$ is the prime bio-marker, being produced almost
exclusively by photosynthesis. The terrestrial atmosphere had a 2-4\%
oxygen abundance since an age of $\sim 2$~Gyr, with a rise to the
present-day levels of 20-30\% starting only after another $\sim
1.5$~Gyr, probably due to the appearance of large vascular land plants
(e.g., Canfield 2005; Goldblatt et al. 2006; Kump et al. 2011).  If
all life on Earth ceased, abiotic processes would remove all oxygen
from the atmosphere within $\sim 10^6$~yr.

Ozone, with a strong signature at 9.6$\mu$m, 
is an indirect bio-marker, produced on Earth via UV
illumination of O$_2$ but it can also form via abiotic
processes, and may be masked by CO$_2$ absorption features
(e.g., von Paris et al. 2011; 2013). 
 Direct detection of O$_2$ in exoplanet
atmospheres has been considered mainly via the rather weak absorption
signal at $1.27\mu$m (e.g., Fujii et al. 2012, but see Palle\' et
al. 2009, where the feature is strengthened by oxygen dimers in 
refraction-enhanced lower-atmosphere spectra), 
and via the O$_2$
absorption bands in the red part of the optical spectrum -- the A, B
and $\gamma$-bands, centered at $\sim 0.76~\mu$m, $0.69~\mu$m, and
$0.63~\mu$m, respectively.  
%The three bands correspond to transitions
%in the O$_2$ molecule from the $\nu=0,1,2$ vibrational levels,
%respectively, in the excited electronic state $b^1\Sigma g^+$, to the
%ground state $X^3\Sigma g^-$. Each vibrational level is split into
%multiple ro-vibrational levels, leading to a rich ``comb'' of lines in
%each band.

Agol (2011) noted that white dwarfs (WDs) have long-lived habitable
zones. If these zones host Earth-mass planets, the planets could
potentially harbour life.
Whereas no planets have yet been detected
around a WD, there is some circumstantial evidence
for their existence. At least one example of a stellar
remnant, a neutron star, has a planetary system (Wolszczan \& Frail
1992).  Sion et al. (2009) estimated, based on the presence of
photospheric metals in at least 15\% of WDs, which possibly results
from the accretion of debris disks, that a similar fraction of WDs
could host terrestrial planets or asteroids. A similar analysis by
Zuckerman et al. (2010) suggested a fraction as high as 30\%. 
Infrared excess emission has been found in several tens of
of such polluted WDs, and has been interpreted as circumstellar dust rings
produced by the tidal disruption of rocky planets or planetesimals,
and subsequently accreted onto the WD, which then displays related
elemental signatures (Zuckerman \& Becklin 1987; Kilic et al. 2005,
2006; Jura et al. 2009; Farihi et al. 2009, 2010ab,2012). 
In several cases (Zuckerman et al. 2007; Klein et al. 2011; Dufour et
al. 2012), the
WD atmospheric abundance pattern has been studied in detail, and found
to be similar to the outer composition
of rocky bodies in the Solar System. Gaensicke et al. (2006, 2007) identified
double-peaked emission from metal-rich gas disks around two WDs, with
the metals thought to result from sublimated solids.

A small planet within several AU cannot survive the asymptotic giant
branch phase, and therefore needs to migrate in from a wider orbit 
to the habitable zone,
after the WD has formed. Even then, Barnes \& Heller (2012) and  Nordhaus \&
Spiegel (2012) have emphasized that the tidal heating of the planet, 
until it had achieved full circularization and synchronization, would
lead to full loss of any water and volatiles present. We note,
however, that the young Earth was also a hot and dry place, but volatiles and
water were then delivered to it by a barrage of comets. The comet impact rate
then decreased to its present low level, greatly lowering
 the biological damage of such impacts. It is not implausible that such 
post-formation 
volatile delivery also could take place on an earth-like planet in a
WD's habitable zone, perhaps driven by the same scattering process that
drove the planet itself to migrate inward after the formation of the WD.
 
Di Stefano et al. (2010), Drake et al. (2010) and Faedi et al. (2011)
noted that, owing to the small sizes of WDs, transits of Earth-sized
and smaller planets orbiting them could easily be detected. As pointed
out by Agol (2011), the habitable zone of a WD is close to its tidal
disruption region. 
At the smallest possible separation, the probability for a transit is
of course the highest, and Agol (2011) has outlined a survey for
discovering such planets.

Here we point out that, compared to planets transiting
main-sequence stars, planets transiting WDs will also
enjoy a much higher contrast of their atmospheric transmission signal
above the background light of their host stars. This makes the
characterization of a planet atmosphere via its transmission spectrum
easily feasible with JWST. In particular, we show that 
the A absorption band of
O$_2$ will be measurable, if O$_2$ is present in the exoplanet at the
level found on Earth over the past 1~Gyr or so. Below,
we calculate the parameters of the required WD survey, and 
the detectability of the sought-after atmospheric
bio-signatures. We then simulate a JWST spectrum of an Earth-like planet
transit accross a WD, and summarize our main
conclusions.

\section{White-dwarf sample requirements}
We begin by estimating the size and type of the WD sample that would 
need to be assembled and surveyed, in order to discover a transiting Earth-like
planet that would be amenable to an atmospheric bio-marker detection. 

White dwarfs follow a narrow mass distribution, with most WDs at
$M_{\rm wd}\sim 0.6 M_\odot$ (Kepler et al. 2007; Kleinman et
al. 2013) or $M_{\rm wd}\sim 0.65 M_\odot$ (Holberg et al. 2008;
Gianninas et al. 2011; Falcon et al. 2012; Giammichele et al. 2012).
Either way, the radius of a WD of such typical mass is $R_{\rm wd}\sim
8500$~km (e.g. Suh \& Mathews 2000). Agol (2011) defines a
``continuously habitable zone'' around a WD that is habitable for at
least 3 Gyr as a WD slowly cools.  For an Earth-density planet
orbiting an $M_{\rm wd}=0.6 M_\odot$ WD, this zone extends from the
tidal disruption limit, at semimajor axis $a\approx 0.005$ AU, out to
$a\approx 0.02$~AU. Adopting $a=0.01$AU as a fiducial value, the
probability for a transit by a planet of radius $R_{\rm_P}$,
\begin{equation}
P_{\rm transit}=\frac{(R_{\rm P}+R_{\rm wd})}{a} 
\end{equation}
obtains a value $P_{\rm transit}=0.01$ for an Earth-like planet with
$R_{\rm_P}=6400$~km. Supposing 20\% of all WDs host an Earth in their
continuously habitable zones, one would need to survey about 500 WDs
to discover one transiting system.
 
Keeping an eye toward discovering the O$_2$ bio-marker in the planet
transmission spectrum, and recalling that bio-generated O$_2$ appeared
on Earth only after $\sim 2$~Gyrs, we will focus on WDs that are $\sim
3$~Gyr old. Although this terracentric bias may be misguided, the
limited available observational resources dictate that we focus on the
systems most likely to have had enough time for oxygen-releasing life
to evolve, rather than finding null results in systems that have not
had enough time. A $0.6 M_\odot$ carbon-oxygen WD that cools for
$3$~Gyr reaches an effective temperature of $T_{\rm eff}\sim 6000$K
(Bergeron et al. 2001)
and a bolometric absolute magnitude $M_{\rm bol}=14.2$. Apart from the
fact that planets around such WDs will have resided in their
 continuous habitable zones
already for a few Gyr, giving potential life some time to evolve, the
temperature at this WD age can also be beneficial in terms of
avoiding a hard UV spectrum that is likely detrimental to life forms.

Ignoring the subtle physics of WD cooling, the rate of change in the
total thermal energy content of a WD equals the power that is
thermally radiated away from its surface, $L\propto T^4$, and the
thermal energy content is proportional to $T$ (assuming a self-similar
temperature profile throughout the cooling history). As a result, the
cooling rate scales as
\begin{equation}
\frac{dT}{dt}\propto T^4 .
\end{equation}
The luminosity function of WDs that are born at a rate $dN/dt$
will be
\begin{equation}
\frac{dN}{dL}=\frac{dN}{dt}\frac{dt}{dT}\frac{dT}{dL}\propto L^{-7/4},
\end{equation}
with the last proportionality assuming a roughly constant
star-formation rate, which leads to a constant WD formation rate,
$dN/dt=$const. The number of WDs per logarithmic luminosity interval
is then
\begin{equation}
\frac{dN}{d(\log L)}\propto L^{-3/4}.
\end{equation}
Harris et al. (2006) and Giammichele et al. (2012) derived consistent
measurements of the luminosity function of local WDs, which is well
fit by a power law with a sharp cutoff at $M_{\rm bol}\approx
15.4$~mag.  The logarithmic slope of their plotted WD luminosity
functions appears remarkably similar to the expected $-3/4$ value
based on the above simple considerations. The cutoff magnitude, for a
typical $0.6M_\odot$ WD, corresponds to a stellar life time of 2.5~Gyr
plus a cooling time of 10~Gyr, that sum to about a Hubble time.  At
$M_{\rm bol}=14.2$~mag, the local density of WDs is $1.6\times 
10^{-3} {\rm
pc}^{-3} {\rm mag}^{-1}$. 
Assembling the brightest (and hence
accessible to JWST spectroscopy)
sample of 500 WDs with $M_{\rm
bol}\sim 14.2$~mag (that are therefore $\sim 3$ ~Gyr old) requires
an all-sky survey for WDs that are within 40 pc. Most of the
WDs in the sample will thus have an apparent magnitude of $\sim 17.2$.
While such an all-sky sample of local WDs does not exist as of yet,
the upcoming {\it
Gaia}\footnote{http://sci.esa.int/science-e/www/area/index.cfm?fareaid=26}
astrometric mission, which will have errors of $<0.1$~milliarcsec at
this magnitude, will easily detect the trigonometric parallaxes of
such WDs and will permit assembling a WD sample such as this. 
 
The WD sample will then need to be monitored photometrically, to find
the one or two cases, if they exist,
 that are transited by habitable-zone Earth-size
planets. As noted
by Agol (2011), about 10\% of any existing transiting planets around these WDs
will be discovered already in {\it Gaia} photometry, so there is a mild
chance to find a transiting earth already at that stage. 
The orbital period will be of order 10 to 20 hours, and the
transit will last of order a minute or two, during which the WD flux will
decrease by order unity. The surveying stage can thus be carried out
with a number of dedicated, 0.5m-class, telescopes, that monitor
each WD continuously with 1-min cadences for a night or two. The
candidate transiting cases can be followed up, using larger
telescopes, with shorter-cadence photometry and radial-velocity
spectroscopic observations, to determine transit ephemerides, planet
sizes, and masses.
 
\section{Bio-marker detection requirements}
 
If and when an Earth-like planet transiting a WD has been discovered, its
atmosphere can be characterized via transmission spectroscopy. When
observing a continuum source through one radial column of the Earth's
atmosphere at resolutions $R\gtrsim 20,000$, the optical O$_2$ lines
become resolved and reach centre-line depths of $\sim 100\%$ in the
A band and $\sim 50$\% in the B band (and considerably weaker in the
$\gamma$ band, which we will not consider further).  At a resolution
of $R\sim 700$ that is practical with JWST in this case, the line
depths are reduced by factors of a few.

Because an Earth-like planet and a WD are of comparable size, the
planet will cover about one-half of the WD disk during a typical
transit. The area of the WD that is covered by the planet
atmosphere will then be about $2\pi R_p n H/3$, where $H$ is the
exponential scale height of the planet atmosphere, and $nH$ is the
height above the planet surface at which a line of sight tangent to
the planet passes a column of atmosphere equal to one vertical column.
This requirement on $n$ can be approximated as
\begin{equation}
\label{5Hequation}
e^{-n} 2\sqrt{n R_p/H}\approx 1 .
\end{equation}     
A planet's radius $R_p$ depends on the planet mass and composition.
The atmospheric scale height $H$ will depend on the gravitational
acceleration at the planet's surface $g$, and thus again on the
planet's mass and composition, but also on the atmospheric composition
and temperature $T_{\rm atm}$,
\begin{equation}
H=\frac{k_BT_{\rm atm}}{\mu g}=\frac{c_s^2}{g},
\end{equation}     
where $k_B$ is the Boltzmann constant, $\mu$ is the mean molecular
mass, and $c_s\equiv (k_BT_{\rm atm}/\mu)^{1/2}$ is the sound speed.
$T_{\rm atm}$, in turn, depends on atmospheric composition and
distance from the parent star.  Thus, planets of differing masses,
compositions, and orbits will have different ratios $R_p/H$, and hence
differing values of $n$ that satisfy the transcendental
equation~(\ref{5Hequation}). Nevertheless, for a range of real planet
$R_p/H$ ratios, $n$ equals approximately 5. We find that for the
Earth, with $R_p=6400$~km and $H=10$~km, $n\approx4.7$. For Jupiter,
with $R_p=70,000$~km and $H=27$~km, $n\approx5.5$. For a hot Jupiter,
with $T_{\rm atm}$ that is 5 times higher than Jupiter's,
$n\approx4.6$. Beckwith (2008) has solved the integral expression for the
atmospheric path length giving $n$, and derived $n=4.2$ for terrestrial
parameters.

The unocculted area of the WD will typically be $\pi R_{\rm wd}^2/2$,
because for the average transit impact parameter roughly one-half of
the WD area will be covered by the planet.  The contrast between the
atmospheric signal $S$ (which we define here as the part of the WD
light that impinges upon the atmospheric annulus) 
and the unocculted background $B$ is then,
\begin{equation}
\frac{S}{B}\approx \frac{4 n H R_p}{3 R_{\rm wd}^2}\approx \frac{1}{170},
\end{equation}
for an Earth-like planet transiting a WD. This is of course a much
larger and more favourable contrast ratio then in the case of earths
transiting main-sequence stars.  The dominant source of noise involves
Poisson fluctuations in the number of photons from the unobscured
light of the background star, in this case the WD.  The required
number of detected signal photons in the transmitted spectrum, for a
given $S/N$, therefore satisfies
\begin{equation}
\frac{S}{\sqrt{170 S}}=S/N, 
\end{equation}
or $S\approx 4000$ signal photons and $B\approx 7\times 10^5$
background photons per spectral resolution element for $S/N=5$.
In \S 4 we show that such a $S/N$ in the continuum of the transmitted 
spectrum is sufficient for detecting clearly the O$_2$ A-band
absorption, which at $R\sim 700$ has a band-center absorption depth of 
about 50\%, and spans several resolution elements. 
 
The half-occulted WD will have about 18 mag,
and hence a photon flux of $f_{\rm ph}\sim 8\times 10^{-5} {\rm
s}^{-1}{\rm cm}^{-2} {\rm \AA}^{-1}$ at a wavelength of $\sim
7000$~\AA.  For JWST, 0.88 of the 6.5-m-diameter collecting area is
unobstructed, and for the Near Infrared Imaging Slitless Spectrograph 
(NIRISS) with the GR700XD grism
($R\sim 700$) the total system throughput at $\sim 7000$~\AA\ is about
0.15 (Doyon 2012).  During a single transit of
length of 2 minutes, about 4,200 photons per 10~\AA\ resolution element
will be accumulated.  Integrating over 160 such 2-minute transits, the
required $S/N$ for detecting the oxygen bands will be obtained.  This
would amount to 5.3 hours of net JWST exposure time (i.e. excluding
telescope overheads), albeit spread out
over daily or twice-daily visits during several months.

If the exoplanet atmosphere has an O$_2$ abundance of
only a few percent, i.e. 1/10 of the present level, as was on Earth
at ages $\sim 2$--$3.5$~Gyr, the O$_2$ signature will be undetectable,
unless one resorts to JWST exposures which are 100 times longer (as
are being considered for studying planets around M stars). On the
other hand,
the near infrared spectrum will be obtained simultaneously in the full
$0.6-2.5~\mu$m NIRISS wavelength range, over which the system
throughput is 30-40\%. Strong signatures of other molecular species in
the the planet atmosphere, such as water and CO$_2$, 
will be detectable with $<1$~hr exposure. 

Naturally, if a suitable habitable-zone Earth-mass planet transiting
a WD is discovered at the survey stage, it will be straightforward to
obtain its atmospheric transmission spectrum with a ground-based
10-m-class telescope, {\it if} the planet atmosphere is significantly
different from the terrestrial example. With a collecting area 3 times
larger than JWST, the total exposure time would be correspondingly
shortened. The added challenge of variable telluric atmospheric
absorption during and out of transit could be addressed by
simultaneously observing, through the instrument slit, a neighbouring
star of similar brightness. However, if the planet atmosphere is also
Earth-like, it will be difficult to detect it from the ground,
after our own highly variable (both temporally and spatially)
atmosphere has imprinted on the spectrum the same signature, but $\sim
170$ times stronger. Indeed, a null ground-based detection of a
planetary atmosphere could mean either that there is no planet
atmosphere or that there is one but it is similar to Earth's, hence
requiring a space-based telescope such as JWST. Alternatively, Snellen
et al. (2013) have described a program for detecting the atmosphere of
an Earth twin utilising the next generation of large telescopes and very
high spectral resolution, to separate the extraterrestrial and
telluric signals by means of the velocity shift of the planet lines.

For planets larger than the WD, which will typically occult the WD
completely during transit, the probability of a transit of the planet
{\it atmosphere} across the WD will be
\begin{equation}
P_{\rm transit-atm}=\frac{2R_{\rm wd}}{\pi a} ,
\end{equation}
i.e., similar to the Earth-like case.  However, given that the planet
temperature will be the same as in the previous Earth-size case
(assuming $a$ in the habitable zone), but the gravity will be larger
in proportion to $R_p$, the scale height $H$ will be proportionally
smaller. The area covered by the planet atmosphere will be typically
just $R_{\rm wd}nH$. For planets larger than the WD, the atmospheric
transmission signal will thus decrease linearly with planet radius,
while the background emission from the WD will remain unchanged, and
hence the required exposure times will increase in proportion to the
planet radius. Among planets transiting WDs, Earth-like planets are
thus optimal for atmospheric transmission studies, since their radius
is similar to that of WDs.
\begin{figure}
%\begin{center}
\includegraphics[angle=-90, width=0.45\textwidth]{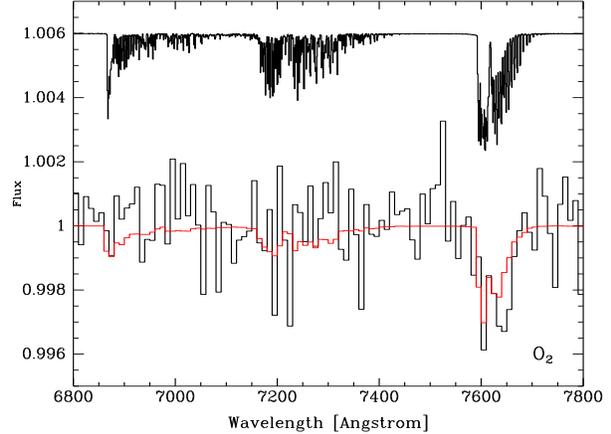}
\caption{Simulated JWST exposure (lasting 5.3~hr in total) of the
transmission spectrum of an Earth-like planet transiting a WD. At the
top, and shifted up arbitrarily by 0.006 for display purposes, 
is a high-resolution
noise-free spectrum, showing the prominent O$_2$ absorption bands. The
red curve is the same spectrum, binned to the JWST NIRISS $R\sim 700$
resolution. The black curve includes the simulated noise. The A-band
of O$_2$ is clearly detectable.}
\label{sdss2dtdia}
%\end{center}
\end{figure}
\begin{figure}
%\begin{center}
\includegraphics[angle=-90, width=0.45\textwidth]{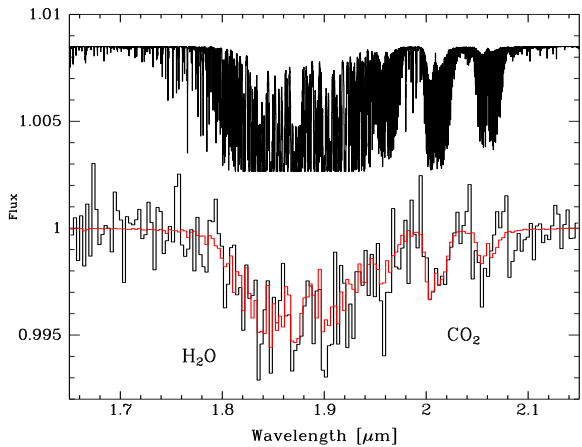}
\caption{Same as in Fig. 1, but for a near-infrared spectral region
dominated by strong absorption features of H$_2$O and CO$_2$. The
H$_2$O feature would be visible also in an exposure shorter by a
factor of 10.}
\label{sdss2dtdia}
%\end{center}
\end{figure}

\section{Spectral simulation}
To demonstrate the feasibility of the proposed observations, we have
simulated the JWST transmission spectrum observation of an Earth-like
planet transiting a WD in the $0.6-2.5~\mu$m wavelength range. The
Earth's empirical atmospheric absorption spectrum, including the
optical O$_2$ bands at $\sim 6000-8000$~\AA, was taken from Wallace et
al. (2011) based on observations of the Sun at different air
masses. For the wavelength band of $0.9-2.5~\mu$m, we use the infrared
atmospheric transmission spectra available at the Gemini
Observatory\footnote{http://www.gemini.edu/?q=node/10789}, which were
calculated using the ATRAN modeling program (Lord
1992). Specifically, we have used the model calculated for 1.0 air
mass at the altitude of Mauna Kea, with a water vapour column of
1.6~mm.

The amount of water vapour one could expect to observe in
transmission in any Earth-like planet, let alone in a planet orbiting
a WD, is highly uncertain. On the one hand, the terrestrial mean water
column is 25~mm, i.e. 15 times larger than we have adopted. On the
other hand, almost all of the terrestrial water vapour is confined
within one scale height, i.e. within a projected atmospheric area that
is 5 times smaller than we have assumed. But then, most or all of this
water-vapour-rich atmospheric layer could be hidden from view in an
exoplanet transmission spectrum by optically thick clouds or haze of
water (e.g. Burgasser 2009).  The uncertainties associated with the
existence and height of clouds as well as the composition and mixing of
the atmosphere, can lead to a very broad range of possible H$_2$O
absorption strengths in an exoplanet atmosphere (e.g. Fujii et
al. 2012).  The 1.6 mm column we
have chosen to illustrate the effect is therefore just one of many
possible values. Refraction in the planet atmosphere is another factor
that could affect the expected signal (Garcia-Munoz et al. 2012). 

We have rebinned the full $0.6-2.5~\mu$m absorption spectrum to the
$R\sim700$ spectral resolution of the NIRISS GR700XD grism, diluted it
with the WD continuum emission (170 times brighter) and added Poisson
noise according to the total expected photon counts, as described in
\S~3. This procedure implicitly assumes that the transmission
spectrum through a height $nH$ is representative of the spectrum at
all heights, from zero to to $nH$ above the surface of the planet. In
reality, for a tangential path below about 10~km, the Earth's
atmosphere becomes quite opaque at most wavelengths, due to absorption
by ozone bands, in addition to water clouds, aerosols, and Rayleigh 
scattering, and therefore these lowest regions will not contribute to the 
transmission signal. On the other hand, at intermediate heights, the 
interesting absorbers actually have a larger optical depth, and hence
layers at those heights contribute a stronger signal, than those at
$nH$ (see Kaltenegger \& Traub 2009). In the balance, the above assumption
provides a reasonable approximation of the expected spectrum. 
   
  Figures~1 and 2 show representative spectral regions of the
simulated spectrum in the optical band around the O$_2$ A and B bands
and in the infrared region around several H$_2$O and CO$_2$
features. The O$_2$ A-band and the H$_2$O and CO$_2$ features are all
clearly detected in this simulated 5.3~hr integrated exposure. We have
verified that the H$_2$O band, which is strong and broad, is
detectable even with $10\%$ of the exposure time.

\section{Conclusions}

We have shown that JWST will be able to detect the O$_2$ bio-signature
in the absorption spectrum of a habitable Earth-mass planet as it
transits a WD, after a total integration time of about 5 hours. Finding
a suitable planetary system requires first assembling a sample of
$\sim 500$ WDs with apparent magnitudes of $\sim 17$ within $\sim
40$ pc, which is feasible with {\it Gaia}, and then monitoring them
with small telescopes in order to find the transits.  
Earth-mass planets in the habitable zones of WDs may
offer the best prospects for detecting bio-signatures within the
coming decade.

\section*{Acknowledgments}
We thank Mukremin Kilic, 
Remko de Kok, Tsevi Mazeh, Dimitar Sasselov, Laura Schaefer, 
Ignas Snellen, Dave Spiegel, Amiel
Sternberg, Lev Tal-Or, and the anonymous referee, 
for useful input.  A. L. acknowledges
support from the Sackler Professorship by Special Appointment at Tel
Aviv University, which enabled this collaboration.  
%The work was also
%supported in part by NSF grant AST-0907890 and NASA grants NNX08AL43G
%and NNA09DB30A (for A.L.). 
D.M acknowledges support by a grant from
the US-Israel Binational Science Foundation.

%\onecolumn
%\newpage

\end{document}